\documentstyle[epsf,aps]{revtex}
\newcommand{\bib}{\bibitem}
\newcommand{\beq}[1]{\begin{equation} \label{#1}}
\newcommand{\eeq}{\end{equation}}
\newcommand{\beqar}[1]{\begin{eqnarray} \label{#1}}
\newcommand{\eeqar}{\end{eqnarray}}
\newcommand{\bra}[1]{\langle {#1}|}
\newcommand{\ket}[1]{|{#1}\rangle}

\begin{document}
\begin{flushright}

Columbia preprint CU--TP--835\\
\end{flushright}
\vspace*{1cm} 
\setcounter{footnote}{1}

\begin{center}
{\Large\bf Non-Forward and Unequal Mass Virtual Compton Scattering}
\\[1cm]
Zhang Chen \footnote{Electronic address: zchen@phys.columbia.edu} \\ ~~ \\
{\it Department of Physics, Columbia University} \\ 
{\it New York, New York 10027} \\ ~~ \\
\end{center}

\begin{abstract} 

We discuss the general operator product expansion of a non-forward
unequal mass virtual compton scattering scattering amplitude. We find 
that the expansion now should be done in double moments with new 
moment variables. There are in the expansion new sets of leading twist 
operators which have overall derivatives, and they mix under 
renormalization. We compute the evolution kernels from which the 
anomalous dimensions for these operators can be extracted. We also 
obtain the lowest order Wilson Coefficients. In the high energy limit 
we find the explicit form of the dominant contributing anomalous 
dimensions and solve the resulting renormalization group equation. We 
find the same high energy behavior as indicated by the conventional 
double leading logarithmic analysis.

\ \\ PACS number(s): 12.38.Bx, 13.60.-r, 11.10.Gh

\end{abstract}

\setcounter{footnote}{0}

\section{Introduction}

Recently there is much interest in the analysis of non-forward scattering 
processes, for example, Deeply Virtual Compton Scattering (DVCS) 
\cite{Ji1,Rad1} and hard diffractive electroproduction of vector 
mesons \cite{Brodsky,Rad2,CFS,Frankf1} in Deeply Inelastic Scattering (DIS). 

Ji \cite{Ji1,Ji2} proposed that one can obtain from DVCS information on 
the Off-Forward Parton Distributions (OFPD) which in this case contain 
new information on long distance physics. He studied their evolution and 
sum rules (in terms of form factors) and obtained certain estimates at low 
energy. Radyushkin also studied the scaling limit of DVCS \cite{Rad1}, and 
generalized the discussion to hard exclusive electroproduction processes 
\cite{Rad2,Rad3}. The non-perturbative information is incorporated in his 
double distributions $F(x,y;t)$ and non-forward distribution functions 
$F_\zeta(X;t)$. He discussed their spectral properties, the evolution 
equations they satisfy, their basic uses and general aspects of 
factorization for hard exclusive processes. A third parametrization for 
the non-perturbative information was proposed by Collins {\it et al}.\ 
\cite{CFS,Frankf1}. They performed a numerical study of their non-diagonal 
parton distributions in leading logarithmic approximation, in which they 
found the nondiagonal gluon distribution $x_2G(x_1,x_2,Q^2)$ can be well 
approximated at small x by the conventional gluon density $xG(x,Q^2)$.

In this paper, we formulate a general operator product expansion 
description for a generic non-forward unequal mass virtue compton 
scattering amplitude(see Fig.\ \ref{ampli}). Because of the non zero 
momentum transfer $r=p-p'$ from the initial state proton to the final 
state proton, the scaling behavior must be different from that in a 
forward case because of the new kinematic degree of freedom 
(the non-forwardness). Indeed we will show that compared with forward 
scattering one has two scaling variables, or equivalently, two moment 
variables $\omega$ and $\nu$ (see (\ref{momvar})). The amplitude should 
now be expanded in terms of double moments with respect to them. There 
are associated with these double moments new sets of operators that have 
overall derivatives in front (see (\ref{operators})), which mix among 
themselves under renormalization via an anomalous dimension matrix. 
The reduced non-forward matrix elements of these operators are thus 
double distribution functions, but they do not seem to have a simple 
probability interpretation. We will focus our attention on the high 
energy behavior of these distribution functions, solving their 
renormalization group equations, and show that in the high energy limit 
the gluonic double distribution function actually reduces to the 
conventional (forward) gluon density and the high energy behavior is
the same as obtained by a conventional double leading logarithmic
(DLL) analysis.

The outline of the paper is as follows. In Section II we define the 
process and the kinematics of the non-forward scattering amplitude 
${\rm T}_{\mu \nu}$. A general tensorial decomposition of 
${\rm T}_{\mu \nu}$ into invariant amplitudes is given. In Section III 
we perform a general operator product expansion of ${\rm T}_{\mu \nu}$,
define new moment variables and give their relationship to the more 
conventional Bjorken type scaling variables. In Section IV we write down 
the renormalization group equations, present the equivalence of 
evolution kernels for the double moments, and calculate explicitly the 
lowest order Wilson Coefficients. In Section V we go to the high energy 
limit to solve the renormalization group equations and make connection 
with the DLL analysis. Section VI gives conclusions and outlook for 
future work.

\section{Decomposition of the Amplitude}

\begin{figure}
\begin{center}
\epsfxsize=3cm
\epsfysize=3cm
\leavevmode
\hbox{ \epsffile{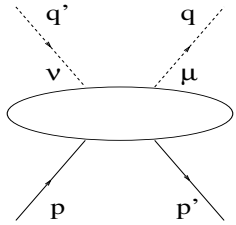}}
\end{center}
\caption{Non-forward Scattering Amplitude}
\label{ampli}
\end{figure}

Consider an unequal mass (non-forward) virtue compton scattering amplitude 
${\rm T}_{\mu \nu}$ as shown in Fig.\ref{ampli}, where the incoming 
photon $q'=q-r$ and the outgoing one $q$ have different invariant masses: 

\beq{masses}
q'^2 = -Q_1^2 , \,\,\,\, q^2 = -Q_2^2 
\eeq 

\noindent We restrict our discussion to amplitudes. Cross sections are
obtained from the square of the amplitudes. In the non-forward case there 
is not a simple optical-theorem relating the imaginary part of an amplitude
to a cross section \cite{Christ}. Our main interest is in the high energy 
limit, so similar to \cite{Brodsky} we take all but one of the components 
of the proton momentum to be zero, same for the momentum transfer $r$.  
Thus $-t=r^2=0$, while $r$ is proportional to the external momentum $p$ and
 
\beq{prop}
r = \zeta p , \,\,\,\,\,\,\,\,\,\,\,\,\,\,\,\, {\rm with} \,\,\,\,\,\,\,\,\,\, 
\zeta = \frac {\,\,\,\,q^2-q'^2} {2 p \cdot q} \,\,\, .  
\eeq

\noindent We can write the spin averaged amplitude, up to an overall 
$\delta$-function in momentum space, as 

\beq{amp1}
{\rm T}_{\mu \nu} (p,q',q) = i \int d^4 z \, e^{-i \bar{q} \cdot z} \, 
\bra{p'} T \, j_{\mu}(-{z \over 2}) j_{\nu} ({z \over 2}) \ket{p}\,
\eeq

\noindent where $ \overline{q}= {1 \over 2} (q'+q)$ and $- \overline{q}^2=
{1 \over 2} (Q_1^2 + Q_2^2) \equiv \overline{Q}^2$ is now the 'natural' 
scale of the scattering process. DIS corresponds to $q=q'$ while 
DVCS corresponds to $q^2=0$. We always suppose $Q_2^2 \leq Q_1^2$ 
so that $\zeta \ge 0$.

Use translation operator one can prove that the conservation of the 
electro$\!-\!$magnetic current now requires the amplitude to satisfy

\beq{conserv}
q^\mu {\rm T}_{\mu \nu} = 0; \,\,\,\,\, q'^\nu {\rm T}_{\mu \nu} = 0. 
\eeq

\noindent ${\rm T}_{\mu \nu}$ is no longer symmetric, rather it is 
invariant under $\mu \leftrightarrow \nu, \,\,\,q \leftrightarrow -q' $. 
Together with current conservation (\ref{conserv}), we obtain the general 
tensorial decomposition of ${\rm T}_{\mu \nu}$ as

\beq{tensordecom}
{1 \over {\sqrt{1-\zeta}}} {\rm T}_{\mu \nu} = \left(-g_{\mu \nu} + 
\frac {q'_\mu q_\nu}{\overline{q}^2} \right){\rm T_1} + 
{1 \over M^2} \left(p_\mu p_\nu - \frac {p \cdot q} {\overline{q}^2}
(p_\mu q_\nu + q'_\mu p_\nu) + ( \frac {p \cdot q} {\overline{q}^2})^2 
q'_\mu q_\nu \right){\rm T_2},
\eeq

\noindent where ${\rm T_i}={\rm T_i}(q'^2, q^2, p \cdot q)$ are invariant 
amplitudes analogous to those in DIS and they are even functions of 
$p \cdot q$. We have pulled out from them an explicit factor of 
$\sqrt{1-\zeta}$ coming from the external spinors as was done in, 
e.g.\ ,\cite{Rad1}. Note we still have only two independent invariant 
amplitudes even in the presence of three invariants, instead of 
two as in the forward case.

\section{Operator Product Expansion}

In the short distance limit, we can perform an OPE for ${\rm T}_{\mu \nu}$ 
as a sum of products of local operators and their corresponding Wilson 
Coefficients \cite{Collins2}: 

\beqar{ope1}
T \, j_\mu (-{z \over 2}) j_\nu ({z \over 2}) 
\stackrel{z_\mu \rightarrow 0}{\longrightarrow} 
\hat{A}_{\mu \nu} \, \sum_{J=1}^\infty
\sum_{n=0}^J \sum_{i=1}^{u_J} \, F_{J,n}^{(i)} (z^2)
\hat{O}_{\mu_1 ... \mu_J}^{(i)(J,n)} (0) {z \over 2}^{\mu_1}
... {z \over 2}^{\mu_J} \nonumber\\ 
+ \hat{B}_{\mu \nu \alpha \beta}
\sum_{J=1}^\infty \sum_{n=0}^J \sum_{i=1}^{u_J} \, E_{J,n}^{(i)}
(z^2) \hat{O}_{\alpha \beta ; \mu_1 ... \mu_J}^{(i)(J,n)} (0)
{z \over 2}^{\mu_1} ... {z \over 2}^{\mu_J},
\eeqar

\noindent where $\hat{A}_{\mu \nu}$ and $\hat{B}_{\mu \nu \alpha \beta}$ are 
conserved tensor structure operators and all indices except $\mu, \nu$ are 
symmetrized as usual. $\hat{B}_{\mu \nu \alpha \beta}$ is the same as in
the forward case, i.e.\ ,

\beq{op1}
\hat{B}_{\mu \nu \alpha \beta} = g_{\mu \alpha}g_{\nu \beta} \Box 
+g_{\mu \nu} \partial_\alpha \partial_\beta
-g_{\mu \alpha} \partial_\nu \partial_\beta
-g_{\nu \beta} \partial_\mu \partial_\alpha \, ,
\eeq

\noindent and corresponds to the tensor structure multiplying ${\rm T}_2$ 
in (\ref{tensordecom}) while we have not found the explicit form of 
$\hat{A}_{\mu \nu}$ which would generate the tensor structure corresponding
to ${\rm T}_1$. But, as in the forward case, once we know ${\rm T}_2$, we 
can obtain ${\rm T}_1$ by a Callen-Gross relationship, at least in
leading logarithmic level (see (\ref{cg})).

Because now the momentum flowing into the local vertex is $r$ instead of 
zero, we have to include in the expansion new sets of operators that 
have overall derivatives. In leading twist (the meaning of which in this 
case will be clear later) the operators for QCD are

\begin{mathletters}
\label{operators}
\begin{eqnarray}
\hat{O}_{\mu_1 ... \mu_J}^{(q)(J,n)} & = & \partial_{\mu_1} ...
\partial_{\mu_n} \, \tilde{q}(0) \gamma_{\mu_{n+1}}i \stackrel
{\leftrightarrow}{\cal D}_{\mu_{n+2}}...i \stackrel{\leftrightarrow}
{\cal D}_{\mu_J}q(0)  \\
\hat{O}_{\mu_1 ... \mu_J}^{(g)(J,n)} & = & \partial_{\mu_1} ...\partial_
{\mu_n} \, F_{\mu_{n+1}}^{\,\,\,\,\,\,\,\,\nu}(0) i \stackrel{\leftrightarrow}
{\cal D}_{\mu_{n+1}}...i \stackrel{\leftrightarrow}{\cal D}_{\mu_J} 
F_{\mu_J \nu}(0).
\eeqar
\end{mathletters}

\noindent After taking the matrix elements between the asymmetric external 
states, and after taking a Fourier transform, the external derivatives would 
eventually be turned into factors of $r_\mu$ while the internal derivatives 
into either $r_\mu$ or $(p+p')_\mu \equiv (2p-r)_\mu$. We thus define the 
two moment variables in the non-forward case as

\beq{momvar}
\omega = \frac {(2p-r) \cdot q} {{\overline{Q}}^2}, \,\,\,\,\,\,\,\,\,\,\,\, 
\nu = \frac {r \cdot q} {{\overline{Q}}^2} \,\, .
\eeq

\noindent The forward case would be the limit $\nu=0$ while DVCS corresponds 
to $\nu=1$.

\begin{figure}
\begin{center}
\epsfxsize=6cm
\epsfysize=3cm
\leavevmode
\hbox{ \epsffile{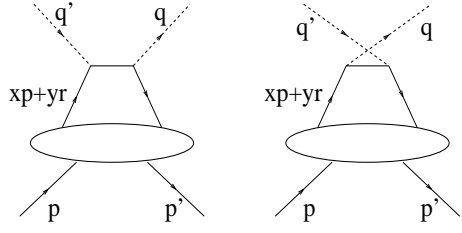}}
\end{center}
\caption{Handbag Diagrams in QCD-Parton Picture}
\label{handbag}
\end{figure}

In a QCD-parton picture, the diagrams contributing to ${\rm T}_{\mu \nu}$
are the so-called 'hand-bag' diagrams as shown in Fig.\ref{handbag}.
If we parametrize the momentum of the scattered parton as $k=xp+yr$ (see,
e.g.\cite{Rad1}), where $x$ and $y$ are two Bjorken type scaling
variables conjugate to $\tilde{\omega} \equiv {{2p \cdot q}
/Q_1^2}$ and $\tilde{\nu} \equiv {{2r \cdot q} / Q_1^2}$, respectively,
our moment variables are related to these quantities via

\beq{relation}
\omega = {Q_1^2 \over \overline{Q}^2} (x \tilde{\omega} + y \tilde{\nu} 
- {1 \over 2} \tilde{\nu}) \,\,\,;\,\,\,\,\,\,\,\,\,\,\,\,\,\,\,\,\,\,\,
\,\,\,\,\, \nu = {Q_1^2 \over \overline{Q}^2} ({1 \over 2} \tilde{\nu}).
\eeq

There will be mixing between operators with same $J$ but different $n$ 
labels. To be more specific, evolution in $\overline{Q}^2$ will lead to 
mixing from $\hat{O}^{(i)(J,n)}$ to $\hat{O}^{(i)(J,n')}$ with 
$0 \leq n \leq n' \leq J$. But we do have the freedom to choose, for 
simplification, at the factorization scale $\mu_0$, that all internal 
derivatives give factors of $(2p-r)$. Thus we can write

\beq{reduce1}
{\bra{p'} \hat{O}_{\mu_1 ... \mu_J}^{(i)(J,n)} \ket{p}}_{(\mu_0)}  = 
r_{\mu_1}...r_{\mu_n}(2p-r)_{\mu_{n+1}}...(2p-r)_{\mu_J} \, {\bra{p'} | 
\hat{O}^{(i)(J,n)} | \ket{p}}_{(\mu_0)}
\eeq

\noindent and the reduced matrix elements will depend only on $J-n$:

\beq{reduce2}
{\bra{p'} | \hat{O}^{(i)(J,n)} | \ket{p}} = {\bra{p'} | \hat{O}^{(i)(J-n,0)} 
| \ket{p}} \,\,\, .
\eeq

Analogous to the forward case, taking the logarithmic derivative of 
$\overline{q}^2$ of the Fourier transformed Wilson Coefficients
$E$ we define the Wilson Coefficients in momentum space 
$\tilde{E}$ as

\beq{wilsonco}
{ {p \cdot q} \over \overline{Q}^2 } i ({1 \over \overline{q}^2})^2
\tilde{E}_{J,n}^{(i)} = ( -i \overline{q}^2 { \partial \over 
{\partial \overline{q}^2}} )^J \int d^4 z \, e^{-i \bar{q} \cdot z}
E_{J,n}^{(i)},
\eeq 

\noindent where we have pulled out an explicit factor of ${{p \cdot q} 
\over \overline{Q}^2}$ to make the form of $\tilde{E}$ simple 
(see (\ref{coeff})). We have

\beqar{ope2}
{\rm T}_{\mu \nu} & = & i^2 \sum_{J,n,i}{\bra{p'}| \hat{O}^{(i)(J,n)} 
|\ket{p}}_{(\mu_0)} \omega^{J\!-\!n} \nu^n
\left( ( -g_{\mu \nu} + \frac {q'_\mu q_\nu} {\overline{q}^2} ) \,
{ {p \! \cdot \! q} \over \overline{Q}^2} \, \tilde{F'}_{J,n}^{(i)}
(\alpha_s, {\overline{Q} \over {\mu_0}}) \right. \nonumber \\
& & \left. + {1 \over {p \! \cdot \! q}} 
\left( p_\mu p_\nu - \frac {p \! \cdot \! q} {\overline{q}^2}
(p_\mu q_\nu + q'_\mu p_\nu) + (\frac {p \cdot q} {\overline{q}^2})^2 
q'_\mu q_\nu \right) \tilde{E}_{J\!-\!2,n}^{(i)}
(\alpha_s,{\overline{Q} \over \mu_0}) \right).
\eeqar

\noindent $\tilde{F'}$ is a combination of $\tilde{E}$ and a similarly
defined $\tilde{F}$. Note we have explicitly indicated their dependence on 
the factorization scale. Non-leading twist terms in this case are suppressed 
by a power of ${1 \over {\overline{Q}^2}}$. Thus we obtain the invariant 
amplitudes as follows:

\begin{mathletters}
\label{invamp}
\begin{eqnarray}
\sqrt{1-\zeta} \, {\rm T}_1 = - \sum_{J,n,i}
{\bra{p'}| \hat{O}^{(i)(J,n)} |\ket{p}}_{(\mu_0)} \omega^{J\!-\!n} \nu^n 
\tilde{F'}_{J,n}^{(i)}(\alpha_s, {\overline{Q} \over \mu_0}) 
\,\,\,\,\,\,\,\,\,\,\,\,\,\,\,\,\,\,\,\,\,\,\,\,\, \\
{ {p \! \cdot \! q} \over M^2} \sqrt{1-\zeta} \, {\rm T}_2 = - \sum_{J,n,i}
{\bra{p'}| \hat{O}^{(i)(J,n)}|\ket{p}}_{(\mu_0)} \omega^{J\!-\!n} \nu^n 
\tilde{E}_{J-2,n}^{(i)}(\alpha_s, {\overline{Q} \over \mu_0}) \,\,\,. \,\,
\,\,\,\,\,\,\,
\eeqar

\end{mathletters}
 
\section{Renormalization Group Equation and Anomalous Dimensions}

\subsection{Renormalization Group Equation}

We concentrate on ${\rm T}_2$. ${\rm T}_1$ can be obtained, as in the case 
of a forward scattering, from the Callen-Gross relationship 

\beq{cg}
{\rm T}_1 = {{p \cdot q} \over {M^2}} {\rm T}_2
\eeq

${\rm T}_2$ can be regarded as a double distribution function and is the 
sum of 'double' moments

\beq{doubmom1}
{\rm T}_2 =  \sum_{J,n} \omega^{J-n} \nu^n {\rm T}_{2}^{(J,n)},
\eeq

\noindent with the double moments ${\rm T}_{2}^{(J,n)}$ as

\beq{doubmom2}
\sqrt{1-\zeta} { {p \! \cdot \! q} \over M^2} {\rm T}^{(J,n)}_
{2,(\overline{Q}, \mu_0)}=- \sum_i {\bra{p'}| \hat{O}^{(i)(J,n)}|\ket{p}}
_{(\mu_0)}\tilde{E}_{J-2,n}^{(i)}(\alpha_s, {\overline{Q} \over \mu_0})
\eeq

\noindent We will eventually analytically continue in $J$, but leave $n$ 
as discrete. The inverse of expansion (\ref{doubmom1}) is

\beqar{inverse}
{\rm T}_{2}^{(J,n)} = \int_c {{d \omega} \over \omega} \int_c {{d \nu} \over
\nu} \, ({1 \over \omega})^{J\!-\!n} ({1 \over \nu})^n {\rm T}_2
(\overline{Q}^2, \omega, \nu) \nonumber \\
= 2i \int_1^\infty {{d \omega} \over \omega} \int_c {{d \nu} \over
\nu} \, ({1 \over \omega})^{J\!-\!n} ({1 \over \nu})^n {\rm W}_2
(\overline{Q}^2, \omega, \nu)
\eeqar

\noindent where $c$ is any contour in the $\omega (\nu)$ plane that encloses 
the origin and ${\rm W}_2= {\rm Im \,T}_2$. Note in obtaining the second 
equality we have used the fact that ${\rm T}_2$ is even in $\omega$. 
${\rm T}_2$ now as a double distribution function is given by

\beq{mellin}
{\rm T}_2(\overline{Q}^2, \omega, \nu) = \int_c {{dJ} \over {2 \pi i}} 
\sum_{n=0}^{\infty} e^{(J-n) \log \omega} \, \nu^n  \, {\rm T}_{2}^{(J,n)}
\eeq

\noindent with the contour $c$ lying parallel to the imaginary axis in the 
$J-n$ plane and to the right of all singularities in that plane.

Under renormalization, these operators scale according to a renormalization
group equation

\beq{rge1}
\mu^2 { d \over {d \mu^2}} \hat{O}^{(i)(J,n)} = \sum_{n'=n}^{J}\sum_{i'}
\tilde{\Gamma}_{nn'}^{ii'} \hat{O}^{(i')(J,n')}
\eeq

\noindent where $\tilde{\Gamma}$ is the anomalous dimension matrix which is 
upper triangular and acts in the product space $J \otimes u_J$. The Wilson 
Coefficients then obey a similar renormalization group equation

\beq{rge2}
\mu^2 { d \over { d \mu^2}} \tilde{E}^{(i)}_{J,n}(\alpha_s(\mu), 
{\overline{Q} \over \mu}) = - \sum_{n'=0}^n \sum_{i'} \tilde{E}^
{(i')}_{J,n'} \,\,\, \tilde{\Gamma}_{n'n}^{i'i}.
\eeq

\noindent We can write the solution as

\beq{solu1}
\tilde{E}^{(i)}_{J,n}(\alpha_s(\mu), {\overline{Q} \over \mu}) = 
\sum_{n'=0}^n \sum_{i'} \tilde{E}^{(i')}_{J,n'}
(\alpha_s(\overline{Q}), 1) \,\,\, {\rm M}_{n'n}^{i'i}(\alpha 
(\overline{Q}), {\overline{Q} \over \mu}),
\eeq

\noindent where ${\rm M}$ is a path ordered exponential of the anomalous 
dimension matrix, formally given as

\beq{expon1}
{\rm M}_{n'n}^{i'i}(\alpha (\overline{Q}), {\overline{Q} \over \mu}) = 
\left( {\cal P} \exp (- \int_{\mu^2}^{\bar{Q}^2} \tilde{\Gamma}(\alpha_s 
(\lambda^2)) { {d \lambda^2} \over \lambda^2}) \right)_{n'n}^{i'i} 
\eeq

\noindent Thus the double moments are given by

\beq{doubmom3}
\sqrt{1-\zeta} { {p \! \cdot \! q} \over M^2}{\rm T}^{(J,n)}_
{2,(\overline{Q}, \mu_0)}=- \sum_{n'=0}^n \sum_{i'i} \tilde{E}_
{J-2,n'}^{(i')}(\alpha_s, {\overline{Q} \over \mu_0}) \,\,\, 
{\rm M}_{n'n}^{i'i}(\alpha (\overline{Q}), {\overline{Q} \over \mu_0}) 
\,\,\, {\bra{p'}| \hat{O}^{(i)(J,n)}|\ket{p}}_{(\mu_0)}.
\eeq 

\subsection{Anomalous Dimensions}

\begin{figure}
\begin{center}
\epsfxsize=10cm
\epsfysize=3cm
\leavevmode
\hbox{ \epsffile{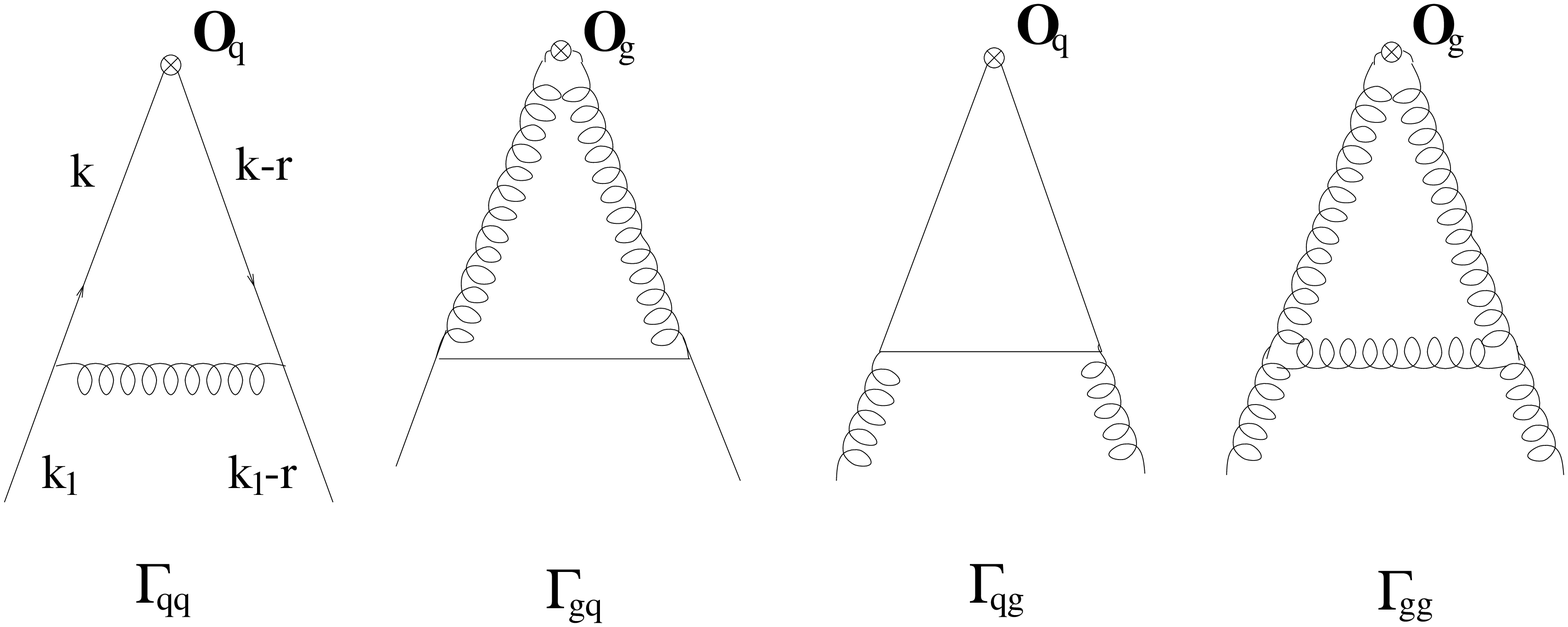}}
\end{center}
\caption{Triangle Graphs for the Anomalous Dimensions}
\label{triangle}
\end{figure}

In light-cone gauge, the lowest order anomalous dimensions of the quark and 
gluon operators are generated by the usual triangle diagrams (note: the
graphs where we have more than two quark/gluon lines meet at the vertex 
are zero because of the gauge choice) as shown in Fig. \ref{triangle}

\noindent Compute the graphs using light-cone variables with the definition

\beq{mtvar}
\omega = { k_+ \over k_{1+}}, \,\,\,\,\, \nu = {r_+ \over k_{1+}},
\eeq

\noindent and the conventions (cf.\ \cite{Mueller1}, also see Appendix A)

\begin{mathletters}
\label{vertex}
\begin{eqnarray}
O_q^{(J,n)(+)}(k,k-r) & = & \gamma_+(2k-r)_+^{J\!-\!n\!-\!1}r_+^n \\
O_g^{(J,n)\alpha \beta}(k,k-r) & = & 2 g_{\alpha \beta} \, n \cdot k 
\, n \cdot {(k-r)} \, (2k-r)_+^{J\!-\!n\!-\!2}r_+^n \,\,\,\,\,\,
\eeqar
\end{mathletters}

\noindent where $n^2=0$ and $n \cdot A = A_+$ for any four vector $A$, 
yields (see Appendix A for details)

\begin{mathletters} \label{adim1}
\begin{eqnarray}
\Gamma_{qq}^{(\!+\!)}\!& = & \!{ \alpha_s \over {2 \pi}} C_F \, \gamma_+ 
k_{1+}^{J-1}\nu^n \, \int_0^1 d \omega \, (2\omega-\nu)^{J\!-\!n\!-\!1}
\left( \Theta(\nu \! - \! \omega) \, {1 \over \nu} \, \omega \, 
(1+ {\nu \over {1-\omega}}) \right. \nonumber \\
& & \,\,\,\,\,\,\,\,\,\,\,\,\,\,\,\,\,\,\,\,\,\,
\left. + \, \Theta( \omega \! - \! \nu ) {1 \over {1\!-\!\nu}} 
{ {1+\omega^2-\nu(1+\omega)} \over {1-\omega}} 
+ \delta(1\!-\!\omega) (-\! 2 I_0 - \log(1\!-\!\nu) + {3 \over 2})
\right) \\
\Gamma_{gq}^{(\!+\!)}\!& = & \!{ \alpha_s \over {2 \pi}} C_F \, \gamma_+ 
k_{1+}^{J-1}\nu^n \! \int_0^1 \! d \omega (2 \!\omega\!-\!\nu)^{J\!-\!n\!-\!2}
\left( \Theta(\nu \! - \! \omega) {1 \over \nu} (\omega^2\!-\!2\omega) 
\!-\!\Theta( \omega \! - \! \nu ){1 \over {1\!-\!\nu}} 
(1\!+\!(1\!-\!\omega)^2\!-\!\nu ) \right), \\
\Gamma_{qg}^{ij} \! &=& \!{ \alpha_s \over {2 \pi}} {1 \over 2} 2g^{ij} 
k_{1\!+}^J\nu^n \! \int_0^1 \! d \omega (2 \! \omega\!-\!\nu)^{J\!-\!n\!-\!1}
\!\left(\!- \!\Theta(\nu \! - \! \omega){1 \over \nu} \omega 
(2\omega\!-\!1\!-\!\nu)\!+\Theta( \omega \! - \! \nu ){1 \over {1\!-\!\nu}} 
(\omega^2 \!+\! (1\!-\!\omega)^2\!-\!\omega \!\nu)\! \right), \\
\Gamma_{gg}^{ij} \!& =& \!{ {\alpha_s C_A} \over {2 \pi}} 2g^{ij} 
k_{1\!+}^J\nu^n \!\int_0^1\!\! d \omega (2\omega\!-\!\nu)^{J\!-\!n\!-\!2}
\!\left(\!\Theta(\nu \! - \! \omega) {1 \over 2} 
({1 \over \nu}(4\omega^3\!\!-\!\omega^2\!\!+\!4\omega)\!+\!
({{\omega^3\!\!+\!\omega^2\!\!-\!2\omega^2\nu} \over {1-\omega}}\!-\!
3(\omega^2\!+\!\omega))) \right. \nonumber \\
& & \left. + { {\Theta( \omega \! - \! \nu )} \over {1\!-\!\nu}} 
({{2(1\!-\!\omega\!+\!\omega^2)^2} \over {1-\omega}}\!+\! \nu \!
{{\nu(1\!+\!\omega^2)\!-\!2(1\!+\!\omega^3)} \over {1-\omega}}) 
\!+\!(1\!-\!\nu) \delta(1\!-\!\omega) ( \!-\!2I_0\!-\!\log(1\!-\!\nu)
\!+\!{ b_0 \over {2C_A}}) \right).
\eeqar
\end{mathletters}

\noindent We have included virtual corrections coming from the 
self-energy graphs that will cancel the colinear singularities at
end point in the $\omega$ integration ($I_0$ terms, see (\ref{diver})
for the definition of $I_0$) and give correct constant terms in
the anomalous dimensions. $\Theta$ is the usual step function and
$b_0= 11- {2 \over 3} n_f$ is the leading coefficient of the QCD 
$\beta$-function. 

Some comments are in order. It is straightforward to show that in the 
forward limit where $\nu=0$, (\ref{adim1}) reduces to the conventional 
Altarelli-Parisi splitting functions \cite{AP}. However, because of the 
non-forwardness, there is no simple probability interpretation for these 
evolution kernels as splitting functions.

The reason we have two terms for each triangle graph is that for different 
$k_+$ integration regions pinching of the $k_-$ pole is different (see 
Appendix A). Thus there is no straightforward optical-theorem type 
dispersion relationship between the cross section and the imaginary part 
of the amplitude. Nonetheless, we can still do analytic continuation in 
$J$ and relate the matrix elements to 'double parton distributions' 
\cite{Christ} which now may not have a direct probability interpretation. 
The second terms of our $\Gamma$ differ from those of \cite{Frankf1} only
because the factor $n \cdot k \, n \cdot (k-r)$ in our (\ref{vertex}b)
is not included in the definition of their gluon vertices.

After we perform the $\omega$ ( i.e.\ $k_+$) integral, we can in principle
obtain the corresponding anomalous dimension matrix in moment space. At first
sight, the $(1-\nu)^{-1}$ factor of the second terms may generate 
a series of infinite sum over powers of $\nu$, which will spoil the locality 
of the vertex thus invalidate the operator product expansion. Detailed 
calculations show that because the lower bound of integration is now $\nu$ 
instead of zero, there will also always be at least one power of $(1-\nu)$ 
coming out from the integral. Furthermore, the highest powers of $\nu$  
cancel completely between the first two terms of each $\Gamma$, leaving the 
highest surviving $\nu$ power as $\nu^{J-1}$ for the quark sector and 
$\nu^J$ for the gluonic sector, as must be the case to make the OPE valid.
All the $I_0$ and $\log(1-\nu)$ terms also cancel completely between real 
and virtual graphs.

Because of the mixing among different $n$ moments, as well as between quark
and gluon sectors, it is very difficult to read off the anomalous dimensions
of the mixing between two operators (with same $J$) of definite $n$ moments.
However, the form of the anomalous dimension matrix simplifies when the
high energy limit is taken, and we will be able to make connection between
this general formalism and the conventional leading logarithmic approximation
(LLA) analysis results.

\subsection{Lowest Order Wilson Coefficients}

The lowest order Wilson Coefficients can be calculated in perturbation
theory from tree level diagrams like the upper part of Fig.{\ \ref{handbag}.
The result is

\beqar{born}
{\rm T}_{\mu \nu}^{(0)} & = &  4 i e_q^2 \sqrt{1-\zeta} \sum_{l=0}^\infty 
\omega^{2l\!+\!1} \left( {{p \! \cdot \! q} \over {\overline{Q}^2}}
( -g_{\mu \nu} + \frac {q'_\mu q_\nu} {\overline{q}^2} ) \right. \nonumber \\
& & \left. + {1 \over {p \cdot q}} \left(p_\mu p_\nu - \frac {p \cdot q} 
{\overline{q}^2}(p_\mu q_\nu + q'_\mu p_\nu) + ( \frac {p \cdot q} 
{\overline{q}^2})^2 q'_\mu q_\nu \right) \right) \, .
\eeqar

\noindent Comparing to (\ref{ope2}) we obtain

\begin{mathletters}
\label{coeff}
\begin{eqnarray}
\tilde{E}^{(q)(0)}_{J,n=0}(\alpha_s(\overline{Q}),1) &=& 4ie_q^2, 
\,\,\,\,\,\,\,\,\, J \,\,\, odd, \, J \geq 3 \,\,\,\,\,\,\,\, \\
\tilde{E}^{(q)(0)}_{J,n}(\alpha_s(\overline{Q}), 1) &=& 0, \,\,\,\,\,
\,\,\,\,\,\,\,\,\,\,\, J,n \,\,\, otherwise. \,\,\,\,\,\,\,\,
\eeqar
\end{mathletters}

\noindent This means at leading order there is no dependence on the
non-forwardness in the Wilson Coefficients.

\section{High Energy Limit}

To exactly solve the renormalization group equations, one needs to 
diagonalize the anomalous dimension matrix in both flavor ($i$) and moment 
($n$) indices. In the high energy limit, however, the situation simplifies.
After analytical continuation in $J$, at high energy, the dominant 
contributions to ${\rm T}_2$ come from the leading (right most) poles in 
the $J$-plane, while in flavor space the gluon anomalous dimension dominates. 
In the terms of the expansion of the path ordered exponential (\ref{expon1}), 
all the factors of the product of anomalous dimensions are 
$\tilde{\Gamma}^{gg}$ except the first one, which should be 
$\tilde{\Gamma^{qg}}$ due to the quark loop, i.e.\, we have to force one 
gluon-quark transition at the end of the evolution. The two flavor
sums in (\ref{doubmom3}) then collapse to $i'=q$ and $i=g$. We evaluate 
{\rm M} between a quark and a gluon state to obtain, again formally, with
$\gamma^{ii'}$ now the anomalous dimension matrix in the moment space,

\beq{expon2}
{\rm M}_{n'n}^{qg}(\alpha (\overline{Q}), {\overline{Q} \over \mu_0}) 
= \left( - \int_{\mu_0^2}^{\bar{Q}^2} {{d \lambda^2} \over \lambda^2}
{\gamma}^{qg}(\alpha_s (\lambda^2)) \right)_{n'l} \left({\cal P}
\exp (- \int_{\mu_0^2}^{\lambda^2} {{d \lambda'^2} \over \lambda'^2}
{\gamma}^{gg}(\alpha_s (\lambda'^2))) \right)_{ln}.
\eeq

\noindent The leading pole terms of the relevant anomalous dimensions 
can be parametrized as (suppressing the $J$ label)

\begin{mathletters}
\beqar{adim2}
\gamma^{qg}_{nn'}(\lambda^2) &=& -{n_f \over {2 \pi}} \alpha_s(\lambda^2)
\,\,\, \tilde{\gamma}^{qg}_{nn'} \,\,\,\,\,\,\,\,\,\,\,\,\,\,\,\, 
{\rm with} \,\,\,\,\,\,\, \tilde{\gamma}^{qg}_{nn'} = 
{{\,\,\,\, 1 + c^{qg}_{n',n}} \over {J-n}} \,\,; \,\, \\
\gamma^{gg}_{nn'}(\lambda^2) &=& -{C_A \over \pi} \alpha_s(\lambda^2)  
\,\,\, \tilde{\gamma}^{gg}_{nn'} \,\,\,\,\,\,\,\,\,\,\,\,\,\,\,
{\rm with} \,\,\,\,\,\,\, \tilde{\gamma}^{gg}_{nn'} = 
{ { 1 + c^{gg}_{n',n}} \over {J-n-1}}\,\, . 
\eeqar
\end{mathletters}

\noindent while $0 \leq n \leq n' \leq J$ for $\tilde{\gamma}_{nn'}$ and 
$c^{i'i}_{n',n}$ vanishes when $n'=n$. Because of the upper triangular 
structure and the leading pole positions of these $\tilde{\gamma}$ matrices, 
we have, since $m \leq l \leq n$,

\beq{diag1}
\sum_l \tilde{\gamma}_{ml} \,\, \tilde{\gamma}_{ln} = \tilde{\gamma}_{mn}
\,\, \tilde{\gamma}_{nn} \equiv \tilde{\gamma}_{mn} \, a_n.
\eeq

\noindent This means in a product of $\tilde{\gamma}$ matrices we can, in 
leading logarithmic approximation (LLA), take only diagonal entries in all 
the factors except the first one. Thus the second factor on right hand side 
of (\ref{expon2}) becomes

\beq{expon3}
{\rm I} + {1 \over a_n^g}(e^{A(\lambda^2)a_n^g} -1) \,\, \tilde{\gamma}^
{gg}_{ln} = {1 \over a_n^g} e^{A(\lambda^2)a_n^g} \,\, \tilde{\gamma}^{gg}
_{ln} \,\,+\,\, {\rm non\!-\!leading terms}.
\eeq

\noindent where $A(\lambda^2)={ {4 C_A} \over b_0} \log \left({{\log 
(\lambda^2 \!/ \Lambda^2)} \over {\log (\mu_0^2 / \Lambda^2)}} \right)$.
Thus, (\ref{expon2}) reduces to

\beq{expon4}
{\rm M}_{n'n}^{qg}(\alpha (\overline{Q}), {\overline{Q} \over \mu_0})
={n_f \over {2 \pi}} \int_{\mu_0^2}^{\bar{Q}^2} { {d \lambda^2}
\over \lambda^2} \alpha_s (\lambda^2) \,\, e^{A(\lambda^2)a_n} \,\,
\tilde{\gamma}^{qg}_{n'n}.
\eeq

\noindent The $n'$ summation in the expression for the double moments 
(\ref{doubmom3}) also collapses in LLA to the diagonal element of 
$\tilde{\gamma}_{qg}$, and we have the final expression as

\beq{doubmom4}
{ {p \! \cdot \! q} \over M^2} \sqrt{1-\zeta} {\rm T}^{(J,n)}_
{2,(\overline{Q}, \mu_0)}=-\tilde{E}_{J-2,n}^{(q)}(\alpha_s, {\overline{Q} 
\over \mu_0}) \,\,\, {\bra{p'}| \hat{O}^{(g)(J,n)}|\ket{p}}_
{(\!\mu_0\!)} \, {n_f \over {2 \pi}} \int_{\mu_0^2}^{\bar{Q}^2} 
{ {d \lambda^2} \over \lambda^2} \alpha_s (\lambda^2) \,\, 
e^{{A(\lambda^2)} \over {J-n-1}} {1 \over {J-n}}.
\eeq

\noindent The invariant amplitude ${\rm T}_2$, after analytically 
continuation in $J$ becomes

\beqar{invamp2}
-{ {p \! \cdot \! q} \over M^2} \sqrt{1-\zeta} {\rm T}_2( \overline{Q}^2,
\omega,\nu) =\int {{dJ} \over {2 \pi i}} \sum_{n=0}^\infty \nu^n  
\int_{\mu_0^2}^{\bar{Q}^2} { {d \lambda^2} \over \lambda^2} 
\alpha_s (\lambda^2) {n_f \over {2 \pi}} \,\,\,\,\,\,\,\,\,\,\,\,\,\,\,\,
\,\,\,\,\,\,\,\,\,\,\,\,\,\,\,\,\,\,\,\,\,\,\,\,\,\,\,\,\,\,\,\,\, \, 
\,\,\,\,\,\,\,\,\,\,\,\,\,\,\,\,\,\,\,\,\,\,\,\,\,\,\,\,\,\, \nonumber \\
\,\,\,\,\,\,\,\,\,\,\,\,\,\,\,\,\,\,\,\,\,\,\,\,\,\, 
\cdot \tilde{E}_{J-2,n}^{(q)}(\alpha_s, 
{\overline{Q} \over \mu_0}) \,\, {\bra{p'}| \hat{O}^{(g)(J,n)}|\ket{p}}_
{(\!\mu_0\!)} {1 \over {J\!-\!n}} \exp \left( (J\!-\!n) \log \omega+
{{A(\lambda^2)} \over {J\!-\!n\!-\!1}} \right).
\eeqar

A saddle point approximation in high energy limit for the $J$ 
integartion leads to

\beqar{invamp3}
-{ {p \! \cdot \! q} \over M^2} \sqrt{1-\zeta}{\rm T}_2 = {n_f \over {2 \pi}} 
\sqrt{ \pi \over {\log^{3/2}\omega}} e^{\log \omega} \int_{\mu_0^2}^
{\bar{Q}^2} { {d \lambda^2} \over \lambda^2} \alpha_s (\lambda^2) 
A(\lambda^2)^{1/4} { 1 \over {1 + \sqrt{ {A(\lambda^2)} 
\over {\log \omega}}}} \,\, \nonumber \\
\,\,\,\, \cdot e^{2 \sqrt{A(\lambda^2) \log \omega}} \,\,\sum_
{n=0}^\infty \nu^n \, \tilde{E}_{J-2,n}^{(q)} (\alpha_s, 
{\overline{Q} \over \mu_0}) 
{\bra{p'}| \hat{O}^{(g)(J,n)}|\ket{p}}_{(\mu_0)} 
\,\,. 
\eeqar

\noindent Because the saddle point fixes the value $J-n=J_s-n=1+ \sqrt{{A 
\over {\log \omega}}}$, which fixes the number of internal derivatives 
inside the gluon operator, recall (\ref{reduce2}) it is clear that the 
reduced matrix element in (\ref{invamp3}) is the conventional (forward) gluon 
distribution function in moment space with a shifted moment label $J_s-n$. 
Despite the formal summation over $n$, which is essentially the only 
difference introduced in this limit by the non-forwardness, only one 
non-perturbative input is needed, which is the gluon density. 
This means that although in general there will be complicated dependence 
on $\nu$ in the double distributions, (in high energy and hard scattering
limit) it is nontheless perturbative. At leading order the $n$ summation
collapses when we take the lowest order Wilson Coefficients given in 
(\ref{coeff}) with the flavor averaged quark charge 
$n_f <e_q^2> =\sum_f e_q^2$ . The final expression of the double 
distribution ${\rm T}_2$ is

\beqar{result}
{{p \! \cdot \! q} \over M^2} \sqrt{1\!-\!\zeta} \,
{\rm T}_2(\overline{Q}^2\!,\omega\!, \nu)\!=\!
{ {i n_f \! <\!e_q^2\!>} \over {2 \sqrt{\pi} b_0 }} \! 
\left(\!{1 \over { { {4C_A} \over b_0 } \log^5 \omega A(\overline{Q}^2)}} 
\!\right)^{1 \over 4}\! \,\, \nonumber \\
\cdot {\bra{p'} | \hat{O}^{(g)(J_s\!-\!n,0)}|\ket{p}}_{(\mu_0)} \,\,\,
e^{\log \! \omega} \,\,\, e^{\! 2 \sqrt{{ {4C_A} \over b_0 } 
A(\bar{Q}^2) \log \! \omega}} .
\eeqar

\noindent We can see clearly that the leading high energy behavior is 
exactly the same as that given by a forward direction leading 
logarithmic analysis. 

\section{Conclusion and Outlook}

We have formulated a general operator product expansion of a non-forward 
and unequal mass virtual compton scattering amplitude. We found that, 
because of the non zero momentum transfer, the expansion now should be 
done in double moments with respect to the moment variables defined 
in (\ref{momvar}). The double moments can be parametrized as products 
of Wilson Coefficients, which can be computed perturbatively, and 
non-forward matrix elements of new sets of quark and gluon operators, 
namely, double distribution functions. These operators mix among 
themselves under renormalization and they obey a renormalization 
group equation. We calculated the equivalence to evolution kernels of 
double distribution from which the anomalous dimension matrix of the 
operators can be extracted.  

In high energy limit, we found the leading contributions of the
anomalous dimension and used them to solve the resulting renormalization 
group equation. We recovered the conventional DLL analysis results and 
made the connection that in fact the double distributions are proportional 
to the conventional forward gluon density in this limit. This justifies
previous analyses using DLL and forward gluon density on non-forward
processes like the exclusive vector meson production (e.g.\ 
\cite{Brodsky}). Furthermore, we have the formalism to proceed in principle 
beyond DLL to next to leading order.

Our analysis is quite general, we can go from DIS (forward case, 
corresponding to $\nu =0$) to DVCS (corresponding to $\nu=1$),
while diffractive vector meson production ($\nu$ close to $1$)
can also be studied by replacing the second quark-photon vertex 
with effectively a wavefunction of the vector meson.

There are still more details of the analysis that need to be filled in. The 
explicit form of the anomalous dimension matrix is not yet written down,
the dispersion relationship between the invariant amplitudes ${\rm T}_i$'s 
and the double structure functions need to be clarified (for example,
the exact physical meaning of the fact that different $k_-$ poles are being
picked out at different $k_+$ integration regions). Some work towards 
these ends is still in progress.

\section*{Acknowledgements}

I want to thank Prof. A.H.\ Mueller for his guidance and advice. Without
his support this work would not have been possible. This research is 
sponsored in part by the U.S.\ Department of Energy under grant 
DE-FG02-94ER-40819.

\appendix

\section{}

% Appendix A

In this appendix we will show in some detail the computation that leads to
(\ref{adim1}), in particular, the quark-quark anomalous dimension.

Using standard Feynman rules and light-cone (LC) gauge we can write down 
the value of the first diagram in Fig. \ref{triangle} as

\beq{gamma1}
\Gamma_{qq}^{+,1} = (ig)^2 C_F \int { {d^4 k} \over { (2 \pi)^4}}
\gamma_\alpha {i \over { \not{k}-\not{r}}} O_q^{(J,n)(+)}(k,r)
{i \over \not{k}} \gamma_\beta { {-i D^{\alpha \beta} (k_1 -k)}
\over { (k_1 -k)^2}}
\eeq

\noindent where the quark vertex $O_q$ is given in (\ref{vertex}) and the
LC gluon projector (numerator of the LC gluon propagator)

\beq{proj}
D^{\alpha \beta}(k) = g^{\alpha \beta} - { {n_\alpha k_\beta + k_\alpha
n_\beta} \over {n \cdot k}}
\eeq

\noindent with $n^2=0$ and $n \cdot A = A_+$ for any four vector $A$. 
Writing the integral of the loop momentum in terms of light-cone 
variables we get

\beq{gamma2}
\Gamma_{qq}^{+,1}= -i { {g^2 C_F} \over { (2 \pi)^4}} \int 
d^2 \underline{k} \, dk_+ dk_- { {(2k-r)_+^{J\!-\!n\!-\!1}r_+^n A(k,r)} 
\over { (k^2 + i \epsilon) ((k-r)^2 + i \epsilon) 
((k_1-k)^2 + i \epsilon)}}
\eeq

\noindent where 

\beq{denom1}
A(k,r)= \gamma_\alpha (\not{k}-\not{r}) \gamma_+ \not{k} \gamma_\beta
D^{\alpha \beta}(k_1-k).
\eeq

We perform the $k_-$ integration first, in terms of the poles of the
integrand. However, the positions of poles in $k_-$ are different when 
$k_+$ is in different regions. More specifically,
when $k_+ < 0$ and $k_+ > k_{1+}$, all the poles in $k_-$ lie
at the same side of the real $k_-$ axis and we can distort the contour
so the integration over $k_-$ gives zero. On the other hand, when 
$0 \leq k_+ \leq r_+$, we can complete the contour in the lower-half-plane
and pick up the pole at $k_-= {{\underline{k}^2} \over {2 k_+}} - i 
\epsilon \equiv k_-^{(1)}$, while when $r_+ \leq k_+ \leq k_{1+}$ we 
complete the contour in the upper-half-plane to pick up the pole at 
$k_-= { {\underline{k}^2} \over {2 (k_+ - k_{1+})}} + i \epsilon \equiv
 k_-^{(2)}$. After evaluating the residues we have

\beqar{gamma3}
\Gamma_{qq}^{+,1}= -i {{\alpha_s C_F} \over { (2 \pi)^2}} 
\int { {d^2 \underline{k}} \over {\underline{k}^4}} \left(
{ { -2 \pi i} \over {2 r_+ k_{1+}}} \int_0^{r_+}
dk_+ k_+ A^{(1)} (2k-r)_+^{J\!-\!n\!-\!1}r_+^n \, \right. 
\,\,\,\,\,\,\,\,\,\,\,\,\,\,\,\,\,\,\,\,\,\,\,\,\,\, \nonumber \\
\,\,\,\,\,\,\,\,\,\,\,\,\,\,\,\,\,\,\,\,\,\,\,\,\,\,\,\,\,\, \left. 
+ { {2 \pi i} \over { -2 k_{1+} (k_1-r)_+}} \int_{r_+}^{k_{1+}}
dk_+ (k_1-k)_+ A^{(2)} (2k-r)_+^{J\!-\!n\!-\!1}r_+^n \right) ,
\eeqar

\noindent where $A^{(1)}$ and $A^{(2)}$ are $A(k,r)$ evaluated at 
the poles $k_-^{(1)}$ and $k_-^{(2)}$, respectively. We are interested
in the logarithmic divergence in the graphs so in computing them we keep
only the leading powers of $\underline{k}$ which is quadratic. We obtain

\begin{mathletters}
\beqar{denom2}
A^{(1)} & = & -2 \underline{k}^2 \gamma_+ (1 + {\nu \over {1 - \omega}})
\,\,\,\,\,\,\,\,\,\,\,\,\,\,\,\,  \\
A^{(2)} & = & - { {2 \underline{k}^2 \gamma_+} \over {(1 - \omega)^2}}
(1 + \omega^2 - \nu (1 + \omega))
\eeqar
\end{mathletters}

\noindent with $\omega$ and $\nu$ as defined in (\ref{mtvar}). Thus we have

\beqar{gamma4}
\Gamma_{qq}^{+,1}  =  { \alpha_s \over {2 \pi}} C_F \int \,
{{d \underline{k}^2} \over \underline{k}^2} \gamma_+ k_{1+}^{J-1}\nu^n \,
\left( {1 \over \nu} \int_0^\nu \, d\omega \, (2\omega-\nu)^{J\!-\!n\!-\!1}
\, \omega \, (1+ {\nu \over {1-\omega}}) \right. 
\,\,\,\,\,\,\,\,\,\,\,\,\,\,\, \nonumber \\
 \left. \,\,\,\,\,\,\,\,\,\,\,\,\,\,  
+ {1 \over {1\!-\!\nu}} \int_\nu^1 \, d\omega \, 
(2\omega-\nu)^{J\!-\!n\!-\!1} \, { {1+\omega^2-\nu(1+\omega)} \over 
{1-\omega}} \right).
\eeqar

\noindent The $\omega$ integration is divergent at $\omega=1$. This 
divergence is cancelled by the self-energy graphs, the value of which
in LC gauge can be readily taken from \cite{Curci} as

\beq{selfe}
Z_F(x)=1+ {{\alpha_s C_F} \over {2 \pi}} { 2 \over \epsilon}
( -2 I_0 - 2 \log{|x|} + {3 \over 2}),
\eeq

\noindent where $I_0$ is the colinear divergence

\beq{diver}
I_0 = \int_0^1 { {dz} \over z} = \int_0^1 { {dz} \over {1-z}}
\eeq

\noindent and $Z_F$ depends on the longitudinal momentum fraction $x$. 
So in our case the self-energy contribution from the $k_1-r$ line 
should have $x= n \cdot (k_1 -r) / n \cdot k_1 = 1 -\nu$ while the
$k_1$ line has simply $x=1$. Adding $ {1 \over 2} ( Z_F(1) + Z_F(1-\nu))$ 
to (\ref{gamma4}) and identifying the logarithmic divergence 
$\int { {d \underline{k}^2} \over { \underline{k}^2}} = 
{ 2 \over \epsilon}$ we will obtain the expression of $\Gamma_{qq}$ as in 
(\ref{adim1}). Note there are no other diagrams contributing in LC gauge.

The computation of the gluonic sector follows similarly. The one point
one needs to pay attention to is the form of the gluon vertex in 
(\ref{vertex}). The full tensorial structure of the LC gluon vertex 
in the non-forward case, generalized from that in \cite{Mueller1}, is  

\beq{gvet1}
V^g_{\alpha \beta}(k,k-r)= g_{\alpha \beta} n \! \cdot \! k 
n \! \cdot \! (k-r) - n \! \cdot \! (k-r) n_\alpha k_\beta
- n \! \cdot \! k (k-r)_\alpha n_\beta + 
k \! \cdot \! (k-r) n_\alpha n_\beta.
\eeq

\noindent This vertex is to be contracted with gluon lines 
$D^{\alpha \alpha'}(k)$ and $D^{\beta \beta'}(k-r)$. It is straightforward
to verify that 

\begin{mathletters}
\beqar{gvet2}
D^{\alpha \alpha'}(k) \, V^g_{\alpha' \beta'}(k, k-r)
\, D^{\beta' \beta}(k-r) \equiv V^g_{\alpha \beta}(k, k-r)  \,\,\,\,\,\, \\
\,\,\,\,\,\, D^{\alpha \alpha'}(k) \, g_{\alpha' \beta'} n \! \cdot \! k 
n \! \cdot \! (k-r) \, D^{\beta' \beta}(k-r) \equiv V^g_{\alpha \beta}(k, k-r),
\eeqar
\end{mathletters}

\noindent thus we can substitute the full gluon vertex by its first term
because of the LC projectors of the two gluon lines connected to 
the vertex and use (\ref{vertex}).

\end{document}